\documentclass[aps, prd]{revtex4-2}
\usepackage{xcolor}
\usepackage{float}
\usepackage{amsmath,amsfonts,amssymb,epsfig,graphicx}
\usepackage{amsmath,mathrsfs}
\usepackage{bbold}
\usepackage{bm}
\usepackage{amsfonts}
\usepackage{amssymb}
\usepackage{slashed}
\usepackage[normalem]{ulem}
\usepackage{lineno}
\usepackage{comment}
\usepackage{afterpage}
\usepackage{titlesec}
\usepackage{booktabs}
\usepackage{multirow}
\usepackage{subcaption} 
\usepackage{setspace}
\usepackage{tikz}
\usepackage[compat=1.1.0]{tikz-feynhand}
\usetikzlibrary{patterns, math}
\usetikzlibrary{positioning}
\usepackage{tabularx}
\usepackage{graphicx}
\usepackage{subcaption} 

\usepackage{hyperref}
\titleformat{\section}{\normalfont\Large\bfseries}{\thesection}{1em}{}
\newcommand{\ket}[1]{\left|#1\right\rangle}
\newcommand{\bra}[1]{\left\langle#1\right|}

\newcommand{\Tr}{\mathrm{Tr}}

\newcommand{\ttb}{t\bar{t}}
\newcommand{\MGVv}{\texttt{MadGraph5\_aMC@NLO}}
\graphicspath{{./figures/}}

\begin{document}

\title{Searching Quantum Entanglement in \(p\ p\to Z\ Z\) process}

\author{Alim \surname{Ruzi}}
\email{alim.ruzi@pku.edu.cn}
\affiliation{State Key Laboratory of Nuclear Physics and Technology, School of Physics, Peking University, Beijing, 100871, China}

\author{Youpeng \surname{Wu}}
\email{youpeng@pku.edu.cn}\thanks{Corresponding author.}
\affiliation{State Key Laboratory of Nuclear Physics and Technology, School of Physics, Peking University, Beijing, 100871, China}

\author{Ran \surname{Ding}}
\email[]{ran.ding@cern.ch }
\affiliation{State Key Laboratory of Nuclear Physics and Technology, School of Physics, Peking University, Beijing, 100871, China}

\author{Qiang \surname{Li}}
\email[]{qliphy0@pku.edu.cn}
\affiliation{State Key Laboratory of Nuclear Physics and Technology, School of Physics, Peking University, Beijing, 100871, China}

\begin{abstract}
Recent studies have shown that observing entangled particle states at a particle collider like Large Hadron Collider (LHC) and testing violation of Bell inequality in them can open up new research area for high energy physics study. We examine the presence of quantum entanglement in the $pp\to ZZ\to 4\ell$ process at leading order. We apply generally recognized method, quantum state tomography, to reconstruct spin density matrix of the joint $ZZ$ system, through which all the relevant observables can be obtained. 
The angular distribution of the final leptons are obtained from simulated events using Monte-Carlo program, which is used to reconstruct spin density matrix. Non-zero value of the lower bound of the concurrence is measured with simulated data.
The numerical analysis shows that with the luminosity corresponding to LHC Run 2+3, entanglement can be probed at $2 \sigma$ level and up to 3.75$\sigma$ level for High-Luminosity LHC data ($3 \rm{ab}^{-1}$), revealing the possibility of finding quantum entanglement in real collider experiment.
\end{abstract}

\maketitle

\section{Introduction} 

Quantum entanglement~\cite{Horodecki:2009zz}, a cornerstone of quantum mechanics, represents a fascinating phenomenon where two or more particles become intricately correlated, such that the change in one particle's quantum state instantaneously influences the state of another termed as "spooky action at a distance" by  Einstein et. al in their paper~\cite{Einstein:1935rr}. This phenomenon might be the most outstanding characteristics of quantum mechanics that truly separates quantum mechanics from classically deterministic theory. This could be proven by testing the violation of Bell inequalities~\cite{Bell:1964kc} using the theory of quantum mechanics, which is unfeasible within any theory that advocates local realism.

Studies about quantum entanglement of multi-particle system at the highest energy frontiers of particle physics has recently  gained enormous interest among both theorists and experimentalists. There have been breakthroughs in the study of quantum entanglement and quantum information science at the high energy frontier~\cite{Afik:2022kwm,Aoude:2025jzc,Afik:2025ejh,Fang:2024ple,Gabrielli:2024kbz, Fabbrichesi:2024rec, Fabbrichesi:2023idl,vonKuk:2025kbv}. For example, several works show that violation of Bell Inequality could be measured in heavy quark system, i.e. $\ttb$, $b\bar{b}$ at the Large Hadron Collider (LHC)~\cite{Afik:2020onf, Afik:2024uif,Fabbrichesi:2021npl,Afik:2022dgh, Severi:2021cnj,Aguilar-Saavedra:2022uye,Aoude:2022imd,Han:2023fci,Dong:2023xiw,Severi:2022qjy,Maltoni:2024csn,Aguilar-Saavedra:2024hwd,Aguilar-Saavedra:2024vpd,Cheng:2024btk}. Considering the heavy mass and very short life time, the $\ttb$ system can be an ideal platform to perform quantum tests at the LHC. Both ATLAS and CMS collaborations~\cite{ATLAS:2023fsd,CMS:2024pts,CMS:2024zkc} have measured entanglement among top quark pairs with high sensitivity. Up to now, search for QE and test Bell nonlocality between other pair of system have continuously gained growing interests at collider experiment.

Quantum state tomography~\cite{White:1999sjn,James:2001klt,Thew:2002fom}, determining the density matrix from an ensemble of measurements, has emerged as a cornerstone technique for reconstructing the quantum state of a system, providing invaluable insights into quantum correlations, coherence, and entanglement~\cite{Popescu:1994kjy,Barr:2024djo,Martens:2017cvj,Bernal:2023jba}. While initially developed within the realm of quantum optics and low-dimensional systems, advancements in experimental and theoretical physics have extended its applicability to high-energy particle physics, particularly in exploring the quantum properties of massive gauge particles produced at particle colliders~\cite{Aguilar-Saavedra:2015yza,Aguilar-Saavedra:2022mpg,Aguilar-Saavedra:2022wam,Ashby-Pickering:2022umy,Barr:2021zcp,Larkoski:2022lmv,Fabbrichesi:2024wcd,Ma:2023yvd,Fedida:2022izl,Fabbrichesi:2023jep,Han:2025ewp,Cheng:2025cuv,Cheng:2024rxi,Han:2024ugl,Goncalves:2025qem,DelGratta:2025qyp}. The advent of high-energy colliders such as LHC and the proposed muon collider opens a promising avenue for probing fundamental aspects of quantum mechanics in previously inaccessible regimes.

The study of massive spin-1 particles, such as the electroweak gauge bosons $WW$ and $ZZ$, is particularly compelling in this context. These particles play a critical role in the Standard Model of particle physics, mediating weak interactions and participating in processes that probe electroweak symmetry breaking~\cite{Glashow:1961tr,Higgs:1964pj,Salam:1964ry,Weinberg:1967tq}. At a muon collider~\cite{AlAli:2021let, Boscolo:2018ytm}, where precise beam properties and high luminosity enable clean experimental conditions, the production of such particles offers an unparalleled opportunity to investigate their quantum properties. Specifically, quantum state tomography of $WW$ and $ZZ$ boson pairs provides a direct method to study their spin correlations, polarization states, and entanglement, enriching our understanding of the quantum nature of the Standard Model.

In this work, we explore quantum properties of $ZZ$ system such as quantum correlation, concurrence and checked separability of this state. This is realized through quantum state tomography of two massive spin-1 particles produced in a 13 TeV proton-proton collision, whose spin density matrix is determined from a simulated events of $pp\to ZZ\to4\ell$. Quantum observables like concurrence are measured to examine whether the $ZZ$ pairs are entangled or separable at such an extremely relativistic environment.

\section{Quantum state tomography and Quantum Entanglement}
\subsection{Spin density matrix for ZZ system}

Spin density matrix is mathematical structure that describes the quantum state of a given system, including both pure and mixed states and encapsulating all the observable related to quantum information such as concurrence~\cite{mintert2004concurrence}, quantum discord~\cite{Ollivier:2001fdq}, quantum magic~\cite{Bravyi:2004isx} and entanglement entropy~\cite{Srednicki:1993im}. In quantum mechanics, state vector $\ket{\Psi}$ is dominantly used to describe the quantum mechanical state of the corresponding system. For pure state, the spin density matrix, $\rho = \ket{\Psi}\bra{\Psi}$ has the equivalent role as $\ket{\Psi}$, while for a mixed state, the spin density matrix is defined as the convex sum 
\begin{align}
    \rho = \sum_i p_i \ket{\psi_i}\bra{\psi_i},
    \label{eq:rhomix}
\end{align}
where \(p_i\) is the classical probability of the \(i\)-th pure state \(\ket{\psi_i}\) and satisfies \(\sum p_i=1\). Density matrix is a positive semi-definite operator, which means $\bra{i}\rho\ket{i}$, where $\ket{i}$ is the base state in the complex Hilbert space $\mathcal{H}$. The total sum probabilities $p_i$ in Eq.~\ref{eq:rhomix} results from the requirement that $\rho$ must have unit trace.

To reconstruct the spin density matrix from experimental or a simulated pseudo data a parametrization of it must be applied. As best know in the spin context of quantum mechanics, the Pauli spin matrices $\sigma_i$ is usually chosen to parametrize the spin density matrix of $s = 1/2$ system, which constitutes qubit. For example, the following form of spin density matrix for a two-qubit system is parametrized with $\sigma_i$:
\begin{equation}
    \rho = \frac14 \bigg[I_2\otimes I_2 +\sum_{i=1}^3 (A_i\sigma_i\otimes I_2 + B_iI_2\otimes \sigma_i) + \sum_{i,j=1}^3 C_{ij} \sigma_i\otimes\sigma_j\bigg],
    \label{eq:gellRho}
\end{equation}
where $A_i$ and $B_i$ are components of the polarization vector of each qubit, and $C_{ij}$ is the entry of the correlation matrix. For any a two-qutrit system, composed of two spin-1 particles, the general spin density matrix can be represented using traceless Gell-Mann matrices~\cite{Gell-Mann:1962yej}:
\begin{align}
\rho=\frac{1}{9}[I_3\otimes I_3]
+\sum_{i=1}^8A_i[T^i\otimes I_3]
+\sum_{i=1}^8B_i[I_3\otimes T^i]
+\sum_{i,j=1}^8 C_{ij}[T^i\otimes T^j],
    \label{eq:densityMatrix}
\end{align}
where \(T^i\) are the \(3\times 3\) Gell-Mann matrices with \(i=1,2,\dots,8\), and \(I_3\) is the 3-dimensional identity matrix. The coefficients \(A_i\), \(B_i\) are spin polarization parameters or the components of the polarization for each qutrit, and \(C_{ij}\) are the spin correlation parameters.  These parameters can be obtained by projecting the spin density matrix on the desired subspace basis as:
\begin{equation}
    \begin{aligned}
        A_i = \frac{1}{6}\text{Tr}\left[\rho T_i \otimes I_3 \right],\quad
        B_i = \frac{1}{6}\text{Tr}\left[\rho I_3 \otimes T_i\right],\quad
        C_{ij} = \frac{1}{4}\text{Tr}\left[\rho T_i \otimes T_j\right].
    \end{aligned}
    \label{eq:ABC}
\end{equation}
Analytically, the obtained coefficients in Eq.~\ref{eq:ABC} are Lorentz invariant which depend only on the energy E, momentum or velocity and the scattering angle in the CM frame. Once these coefficients are known, it is quite straightforward to determine the observables quantifying the entanglement in the massive gauge boson system.

\subsection{Entanglement observables}
Quantifying entanglement of the state of a quantum system is generally challenging as the complexity of the problem increases with the system dimensionality. For a pure state described by a single vector in the Hilbert space, or equivalently by a density matrix, this can be solved by their Schmidt decomposition~\cite{Horodecki:2009zz}. As for the mixed states, a general approach is to compute the so-called entanglement witness, a quantities that give sufficient conditions to establish the presence of entanglement in the system. Such an observable quantity is concurrence, which already mentioned in the above section. This is reliable entanglement measure for bipartite system, consisting of two particles~\cite{Bennett:1996gf,Wootters:1997id}.

Usually for a pure state described by a joint-state vector, the corresponding concurrence is defined as
\begin{align}
  \mathcal{C}[\ket{\psi}]=\sqrt{2(1-\Tr[(\rho_r)^2])},
    \label{eq:Cpure}
\end{align}
where \(\rho_r\) is the reduced density matrix of the subsystem obtained by tracing over the degree of freedom of either subsystem. However, concurrence of any mixed state described Eq.~\ref{eq:rhomix} is given by means of the concurrence of the pure states as 
\begin{equation}
    \mathcal{C}(\rho)=\inf_{\{|\psi_i\rangle\}} \sum_i p_i \mathcal{C}\left(\left|\psi_i\right\rangle\right),
    \label{eq:Cmix}
\end{equation}
where infimum is taken over all the possible decompositions of $\rho$ into pure states. A vanishing value of Eq.~\ref{eq:Cpure} implies that the pure state of the bipartite system is separable. However, due to the complexity in the evolution of mixed states, evaluating its concurrence, Eq.~\ref{eq:Cmix}, to a specific value is quite challenging. Therefore, finding a lower bound of concurrence for mixed states, rather than obtaining the exact value of, also unequivocally indicates the presence of entanglement. This lower bound is analytically computable~\cite{mintert2007observable}:
\begin{align}
   [\mathcal{C}(\rho)]^2 =\mathscr{C}_2[\rho] = 2 \max\left(0,\Tr[\rho^2]-\Tr[(\rho_A)^2],\Tr[\rho^2]-\Tr[(\rho_B)^2] \right),
    \label{eq:C2}
\end{align}
or equivalently, the above lower bound can be written for two-qutrit system of spin-1 massive bosons
\begin{equation}
    \mathscr{C}_2[\rho] = 2\Tr[\rho]^2 - \Tr[\rho^2_A] - \Tr[\rho^2_B] \equiv c^2_{\text{MB}}.
\end{equation}
By using the explicit form of the spin density matrix parametrized as Eq.~\ref{eq:gellRho}, the lower bound of concurrence  $\mathscr{C}_2$ can be written in terms of  the coefficients given in Eq.~\ref{eq:ABC}:
\begin{equation}
    \label{eq:c2_calc}
    c^2_{\text{MB}}=-\frac{4}{9}-6\sum_i A_i^2 - 6\sum_i B_i^2 + 8\sum_{ij}C_{ij}^2.
\end{equation}
The state is entangled if the above quantity is positive, while that lower bound is negative or equal to zero the concurrence test is inconclusive.
\subsection{Extraction of Spin Density matrix from data}
Due to its short lifetime, $Z$ boson can not be detected directly inside the detector. The most abundant decay channel is hadronic decay, which accounts for up to $70\%$ of the total decay. The leptonic decay channel is convenient ot manipulate both in the simulation and in the actual experiments. The information about entanglement of the states are passed to the decay products namely, charged leptons. The angular direction of each of these leptons is correlated with the direction of the spin of their parent $Z$ boson in such a way that the normalized differential cross-section of the process $ZZ\to \ell^+_1\ell^-_1 \ell^+_2\ell^-_2$ may be written as~\cite{Rahaman:2021fcz}
\begin{equation}
    \frac{1}{\sigma} \frac{\mathrm{d}\sigma}{\mathrm{d}\Omega^+ \mathrm{d}\Omega^-} = \left( \frac{3}{4\pi} \right)^2 \operatorname{Tr} \left[ \rho_{V_1 V_2} \left( \Gamma_1 \otimes \Gamma_2 \right) \right]
    \label{eq:sigma}
\end{equation}
in which the solid angle measure $\mathrm{d}\Omega^{\pm} = \sin\theta^{\pm}\mathrm{d}\theta^{\pm}\mathrm{d}\phi^{\pm}$ are written in terms of the spherical coordinates for the momenta of the final charged leptons in the respective rest frame of the decaying particles. The density matrix $\rho_{V_1V_2}$ encodes the production information of the joint $ZZ$ system, while the decay density matrix $\Gamma_i$ passes the relevant entanglement information about the $ZZ$ state to the final decay products, so that we can recover the spin density matrix from these angular distribution. The explicit form of the decay density matrix $\Gamma_i$ can be obtained using polarized decay amplitudes~\cite{Rahaman:2021fcz}
\begin{equation}
    \Gamma = \frac{1}{4}
\begin{pmatrix}
1 + \cos^2\theta - 2 \eta_\ell \cos\theta & 
\frac{1}{\sqrt{2}} (\sin 2\theta - 2 \eta_\ell \sin\theta) e^{i\varphi} & 
(1 - \cos^2\theta) e^{i 2\varphi} \\
\frac{1}{\sqrt{2}} (\sin 2\theta - 2 \eta_\ell \sin\theta) e^{-i\varphi} & 
2 \sin^2\theta & 
- \frac{1}{\sqrt{2}} (\sin 2\theta + 2 \eta_\ell \sin\theta) e^{i\varphi} \\
(1 - \cos^2\theta) e^{-i 2\varphi} & 
- \frac{1}{\sqrt{2}} (\sin 2\theta + 2 \eta_\ell \sin\theta) e^{-i\varphi} & 
1 + \cos^2\theta - 2 \eta_\ell \cos\theta
\end{pmatrix},
\end{equation}
where $\theta$ and $\phi$ are the polar angles of the momentum of the negative charged lepton in the rest frame of mother particle, which is $Z$ boson. The factor $\eta_{\ell}$ is function of electroweak mixing angle~\cite{Aguilar-Saavedra:2017zkn} and is equal to 0.15 when using the latest measurement of mixing angle~\cite{ParticleDataGroup:2024cfk}.

The functions given in Eq.~\ref{eq:Psymbol} and together with the matrix $A_i^j$~\ref{eq:amatrix} in Appendix can be used to extract the polarization and correlation coefficients of the density matrix given in Eq.~\ref{eq:densityMatrix}:
\begin{eqnarray}
    \begin{aligned}
        \hat{A}_i & = \frac{1}{6} \left\langle \tilde{p}_i^1(\hat{\mathbf{n}}_1) \right\rangle, \quad 
        \hat{B}_i  = \frac{1}{6} \left\langle \tilde{p}_i^2(\hat{\mathbf{n}}_2) \right\rangle, \quad
        \hat{C}_{ij} = \frac{1}{4} \left\langle \tilde{p}_i^1(\hat{\mathbf{n}}_1) p_j^2(\hat{\mathbf{n}}_2) \right \rangle,
    \end{aligned}
    \label{eq:abc}
\end{eqnarray}
where $\tilde{p}$ is the corresponding Wigner P symbols for $Z$ boson decay, and given as 
\begin{equation}
    \tilde{p}_i= \sum_{j=1}^8 A_i^jp_j^+.
\end{equation}
The unit vectors $\hat{\mathbf{n}}(\theta, \phi)$ in Eq.~\ref{eq:abc} is defined to point the momentum of the decay leptons in the rest frame.
Because the two $Z$ bosons are indistinguishable, a symmetry factor should be implied by exchanging the labels $i$ and $j$, so that the above coefficients are determined as:
\begin{eqnarray}
    \begin{aligned}
        \hat{A}_i &=\hat{B}_i= \frac{1}{12} \left\langle p_i^1(\hat{\mathbf{n}}_1)+p_i^2(\hat{\mathbf{n}}_2)  \right\rangle, \quad \\
        \hat{C}_{ij}& = \frac{1}{8} \left\langle p_i^1(\hat{\mathbf{n}}_1) p_j^2(\hat{\mathbf{n}}_2)  + p_j^1(\hat{\mathbf{n}}_1) p_i^2(\hat{\mathbf{n}}_2)\right \rangle .
    \end{aligned}
    \label{eq:abc2}
\end{eqnarray}
Eq.~\ref{eq:abc2} provide the means to reconstruct the correlation functions fo the density matrix from the distribution of the lepton momenta and thus allow to derive the expectation value of the lower bound of concurrence $\mathscr{C}_2$ from the data.

\section{Simulation and Numerical results}

\begin{figure}[!htb]
    \centering
    \scalebox{0.8}{
\begin{tikzpicture}
    \begin{feynhand}
        \vertex (f)     at  (-2,2) {\(f\)};
        \vertex (fbar)  at  (-2,-2) {\(\bar{f}\)};
        \vertex (ffz)     at  (0,1);
        \vertex (ffbarz)  at  (0,-1);
        \vertex (z1)     at  (2,1); 
        \vertex (z2)  at  (2,-1);
        \vertex (l1)    at  (4,1.5) {\(\ell\)};
        \vertex (l2)    at  (4,0.5) {\(\bar{\ell}\)};
        \vertex (l3)    at  (4,-0.5) {\(\ell\)};
        \vertex (l4)    at  (4,-1.5) {\(\bar{\ell}\)};

        \propag[fer] (f) to (ffz);
        \propag[fer] (ffz) to (ffbarz);
        \propag[fer] (ffbarz) to (fbar);
        \propag[bos] (ffz) to [edge label=\(Z\)](z1);
        \propag[bos] (ffbarz) to [edge label=\(Z\)](z2);
        \propag[fer] (z1) to (l1);
        \propag[antfer] (z1) to (l2);
        \propag[fer] (z2) to (l3);
        \propag[antfer] (z2) to (l4);
    \end{feynhand}
\end{tikzpicture}
}
\hspace*{3em}
\scalebox{0.8}{
\begin{tikzpicture}
    \begin{feynhand}
        \vertex (f)     at  (-2,2) {\(f\)};
        \vertex (fbar)  at  (-2,-2) {\(\bar{f}\)};
        \vertex (ffz)     at  (-0.5,1);
        \vertex (ffbarz)  at  (-0.5,-1);
        \vertex (z1)    at  (2,1.5); 
        \vertex (z2)    at  (2,-1.5);
        \vertex (l1)    at  (4,2) {\(\ell\)};
        \vertex (l2)    at  (4,1) {\(\bar{\ell}\)};
        \vertex (l3)    at  (4,-1) {\(\ell\)};
        \vertex (l4)    at  (4,-2) {\(\bar{\ell}\)};

        \propag[fer] (f) to (ffz);
        \propag[fer] (ffz) to (ffbarz);
        \propag[fer] (ffbarz) to (fbar);
        \propag[bos] (ffz) to [edge label=\(Z\)](z2);
        \propag[bos] (ffbarz) to [edge label=\(Z\)](z1);
        \propag[fer] (z1) to (l1);
        \propag[antfer] (z1) to (l2);
        \propag[fer] (z2) to (l3);
        \propag[antfer] (z2) to (l4);
    \end{feynhand}
\end{tikzpicture}
}
    \caption{The Feynman diagram of the $pp\to ZZ$ process.}
    \label{fig:signal}
\end{figure}
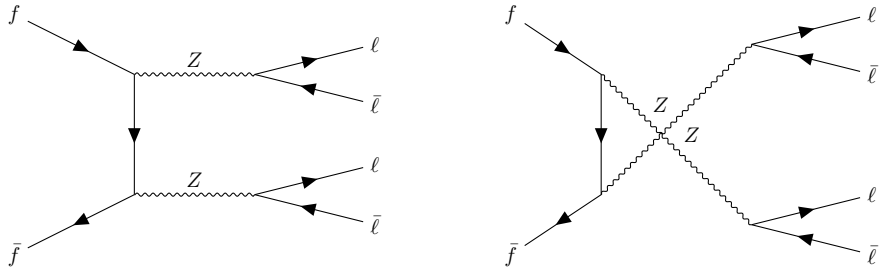
A sensitivity test for the entanglement content is done by conducting pseudo-experiments of the process $pp\to ZZ$ process at the LHC with a center-of-mass energy of \(\sqrt{s}=13\) TeV. Unlike massive gauge boson pair production in the Higgs decay, in which one of the gauge boson is produced off-shell, here both \(ZZ\) bosons are produced on-shell. The Feynman diagram of the process $pp\to ZZ$ process, in which $Z\to\ell^+\ell^-$ $(\ell = e, \mu )$, where $\ell$ can be either electron or muon, is shown in Fig.~\ref{fig:signal}. This is the signal process in our study. Simulation is achieved through publicly available Monte-Carlo software \MGVv ~\cite{alwall2014automated}. During the simulation the relevant spin-correlation and Breit-Wigner effects are include through \texttt{MadSpin} program~\cite{Artoisenet:2012st,Frixione:2007zp} that comes with \MGVv. 

The number of generated events and corresponding cross section sections are given in the Tab.~\ref{tab:processes}. The relevant cross sections are computed using \MGVv at the Leading Order (LO) level. The center-of-mass energy (CM) is set to $\sqrt{s} = 13$ TeV for three different collider luminosities corresponding to the LHC Run 2+3 and HL-LHC luminosities, respectively. When calculating the cross section for both the signal and background process, we use the default Parton Distribution Functions (PDFs) set \texttt{NNPDF}23~\cite{Ball:2012cx} and set the factorization and normalization scale to the physical mass of the $Z$ boson.

\begin{figure}[!htbp]
    \centering
    \includegraphics[width=0.45\linewidth]{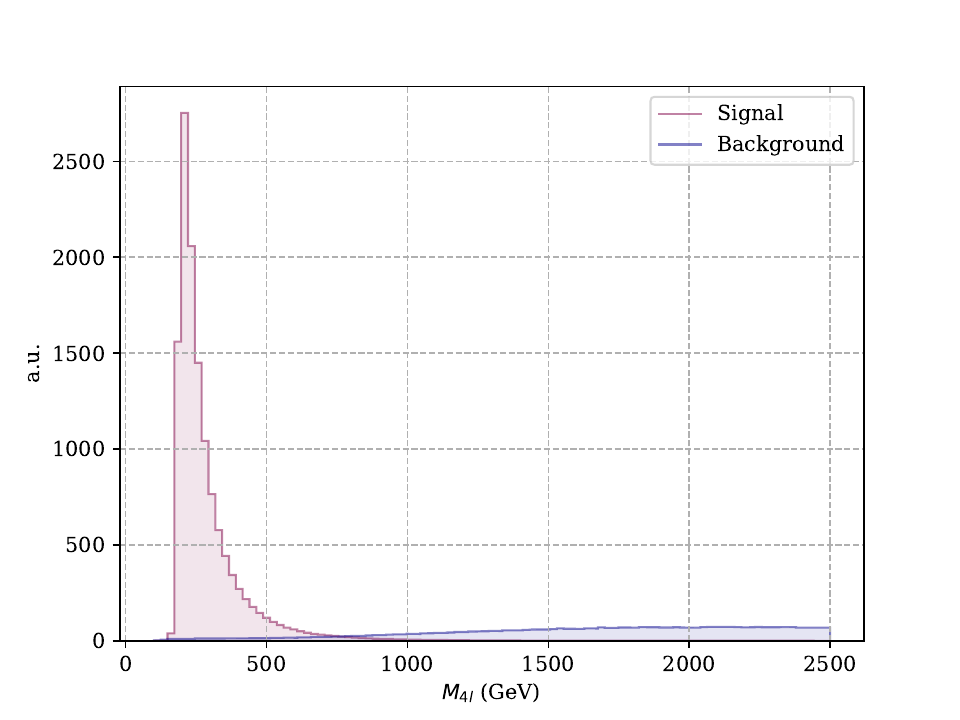}
    \caption{The event distribution of the signal and background process.}
    \label{fig:events}
\end{figure}

\begin{figure}[!htbp]
    \centering
    \includegraphics[width=0.45\linewidth]{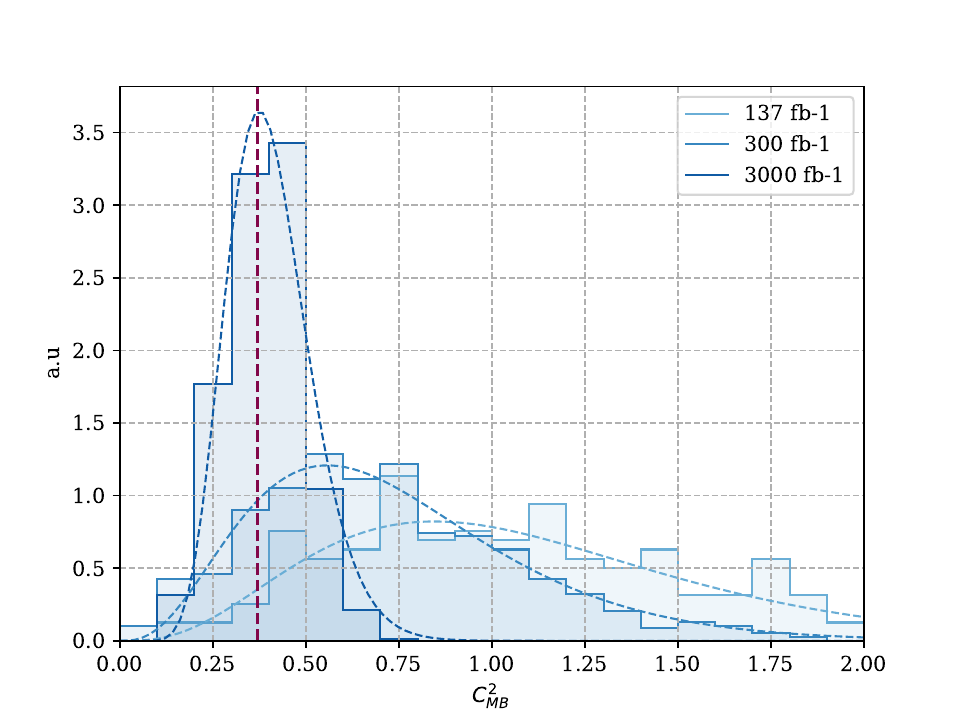}
    \caption{\raggedright $C_{\rm{MB}}^2$ distribution computed from LHC Run 2+3 and HL-LHC luminosity data. Red dashed line is the mean value obtained from HL-LHC luminosity.}
    \label{fig:c2gaus}
\end{figure}

As for the possible background process, we consider the following three process as background to our signal:
\begin{itemize}
    \item $pp\to WWZ$
    \item $pp\to WZZ$
    \item $pp\to ZZZ$
\end{itemize}
\begin{table}[!htb]
    \centering
    \begin{tabularx}{\textwidth}{c|c|c|c|c}
        \toprule
        Process & Cross section (fb) & Events(\(\textrm{Lumi}=137 \rm{fb}^{-1}\)) & Events(\(\textrm{Lumi}=300 \rm{fb}^{-1}\)) & Events(\(\textrm{Lumi}=3 \rm{ab}^{-1}\)) \\
        \midrule
        \(p\ p\to ZZ\)(signal) & \(42.29\) & 5793 & 12687 & 126870 \\
        \(ZZZ\) & \(24.16\) & 3309 & 7248 & 72480 \\
        \(WWZ\) & \(29.76\) & 4077 & 8928 & 89280 \\
        \bottomrule
    \end{tabularx}
    \caption{\raggedright The signal and background processes considered in this work in the first column, total cross section in the second column and the event numbers given for a total luminosity.}
    \label{tab:processes}
\end{table}
Fig.~\ref{fig:events} depicts the event distributions for both the signal and total backgrounds as function of invariant mass of final four leptons. As is shown, the background events does not matter in the signal region. Based on this, we can safely ignore any contamination coming from background events. The \(p p \to Z Z\) signal process exhibits strong background suppression due to the reconstruction of the \(ZZ\) pair via isolated leptons, resulting in minimal background contamination. The dominant background, \(WWZ\), involves two \(W\) bosons decaying to charged lepton and corresponding neutrino, leading to a broad four-lepton invariant mass (\(M_{4\ell}\)) distribution. This distinct kinematic feature allows clear separation from the narrow resonance of the \(ZZ\) signal.  The final lepton coming from Z boson pairs can be identified as two electrons and two muons, four electrons or four muons. For two electrons and two muons, it is easy to distinguish their parent $Z$ boson . In the case where the \(ZZ\) system decays into four same-flavor leptons, we distinguish the signal by combinations: we iterate through all possible lepton pairings and select the combination in which  the invariant mass of the reconstructed Z boson  is closest to the on-shell mass. This strategy is effective because both Z bosons in the signal process are on-shell. 
In steps of the numerical calculation, we extract the polarization and correlation coefficients of the spin density matrix from each single event using Wigner P symbols. Running this procedure over all events gives an average value and the standard deviation of our observables. We produced one million events in total and divide them into 1000 pseudo experiments, each one includes events that matches to the total reachable events in an actual experiments. The statistical uncertainty is determined with these pseudo experiments by repeating over all of the pseudo experiment.
\begin{figure}[!htb]
    \centering
    \includegraphics[width=0.45\linewidth]{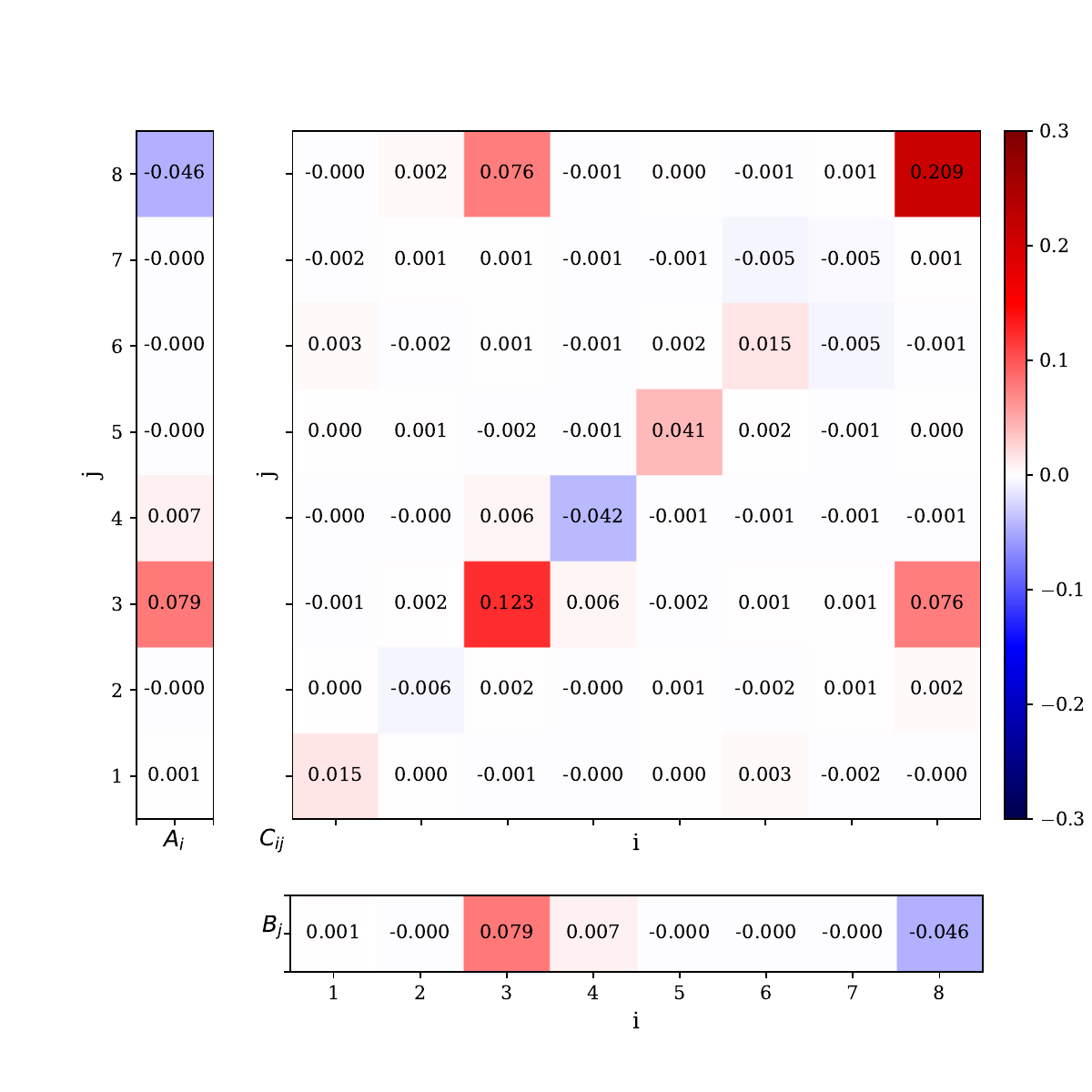}
    \caption{\raggedright The correlation matrix element extracted from the simulated data of $ZZ$ production. The leftmost column contains the $A_i$ parameters of $Z$ boson and the bottom row contains the parameters $B_i$ and the center middle table plot contains $C_{ij}$ coefficients.}
    \label{fig:Cij}
\end{figure}
The parameters of the spin density matrix, $A_i$, $B_i$ and $C_{ij}$  are shown in a matrix plot depicted in Fig.~\ref{fig:Cij}. It is clear that there are large portion of non-zero coefficients indicating the correlations between two $Z$ boson.
\begin{figure}[!htbp]
    \centering
        \includegraphics[width=0.45\textwidth]{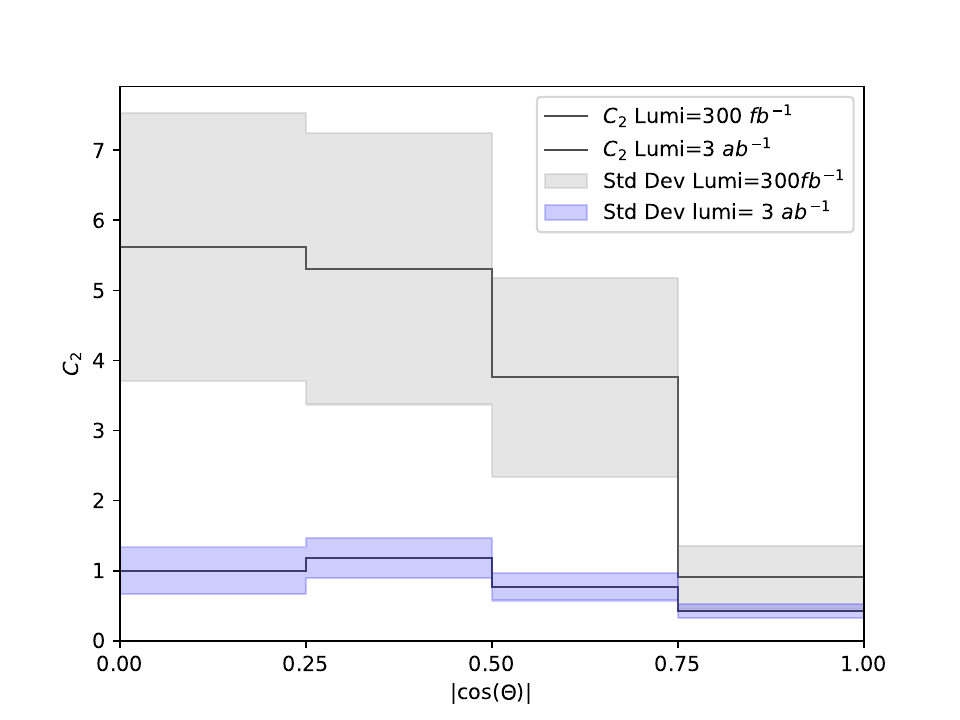}
    \caption{\raggedright Concurrence distribution as function of scattering angle cosine $\cos{\theta}$. The thick central lines in are the central value and colored bands represent statistical deviations obtained from events corresponding to LHC Run 2+3 and HL-LHC luminosities.}
    \label{fig:c2dis}
\end{figure}
The concurrence observable that quantifies entanglement is given as the function of $\cos{\theta}$ which is depicted in Fig.~\ref{fig:c2dis}. For the Run 2+3 data, the gray band, the computed lower bound of the concurrence is relatively larger than one, which is unphysical, while it is close to unity when $\theta\to 0$ or $\pi$. However, the HL-LHC data, the colored one, gives rather physical results when the scattering aligns with beam direction. Because of the high event numbers, the uncertainty of the concurrence for this run is also quite small. Quantum entanglement test is of highly statistical meaning, which means the results computed for the observables at hand, both the mean and deviation, are dependent on the event numbers at the actual experiment. To confirm this statistical nature of the entanglement, a Gaussian distribution of the lower bound of the concurrence, $C_{\rm{MB}}^2$, is obtained corresponding to LHC Run 2+3 and HL-LHC luminosities. The results are shown in the Fig.~\ref{fig:c2gaus}. It can be seen clearly that, $C_{\rm{MB}}^2$ distribution converges to the mean value depicted by the red-dashed line for the HL-LHC data, while it diverges for the Run 2+3 data.

\begin{figure}[!htbp]
    \centering
    \includegraphics[width=0.44\linewidth]{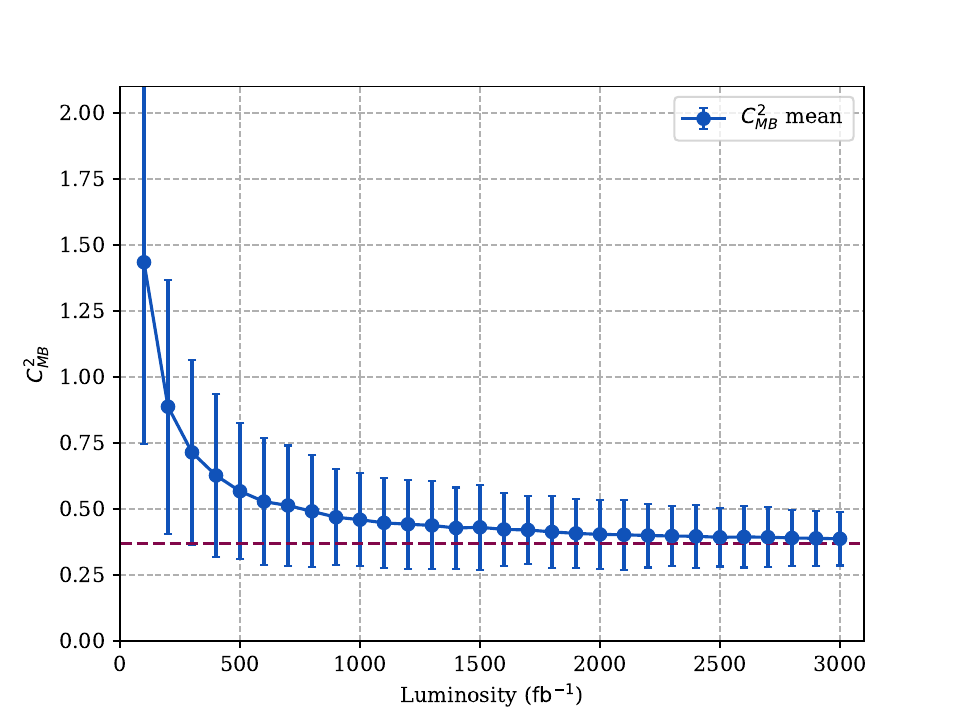}
    \includegraphics[width=0.4\linewidth]{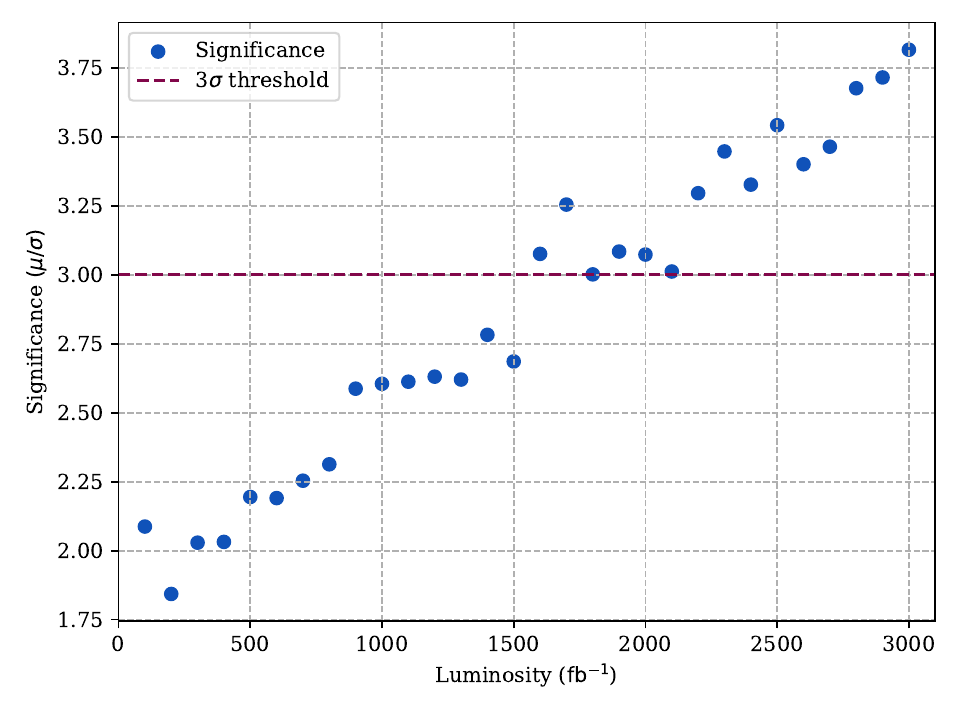}
    \caption{The expectation value (left panel) of $C_{\rm{MB}}^2$ as function of collider luminosity, and the significance distribution (right panel).}
    \label{fig:significance}
\end{figure}

Based on the distributed results shown in Fig.~\ref{fig:c2gaus}, The expectation value of $C_{\rm{MB}}^2$ observable converges to some point when the luminosity of the collider increases. As a test, we compute a luminosity spectrum for $C_{\rm{MB}}^2$ and the results are shown on the left panel of the Fig.~\ref{fig:significance}: the points are the expectation value and the error bar is the uncertainty. Obviously, $C_{\rm{MB}}^2$ converges approximately to 0.375 as the luminosity increases. Meanwhile, the corresponding significance can reach to 3.75 $\sigma$ for the HL-LHC luminosity, which is shown on the right panel fo the Fig.~\ref{fig:significance}. This is well-above the 3$\sigma$ confidence level.

\section{Summary}
Quantum entanglement is one of the most distinctive features of the quantum mechanics. It is widely being examined and tested by particle physics community especially the high energy physics frontiers such as at the LHC and future colliders. In this work, we apply quantum state tomography to $ZZ$ system to inspect the presence of the quantum entanglement in the $pp\to ZZ\to 4\ell$ process. The theoretical framework in which how the spin density matrix is computed or parametrized for the $ZZ$ system is given as well as a through explanation of the quantum state tomography is also provided.

Importantly, a fast Monte-Carlo simulation of the $pp\to ZZ$ is done using \MGVv software so that a large number of pseudo experiments can be used to determine whether quantum entanglement exist or not in the $ZZ$ state.  To achieve this, firstly, Gell-Mann spin matrices are used to parameterize the spin density matrix. Secondly, one million events corresponding to LHC Run 2+3 and HL-LHC luminosity are produced for $pp\to ZZ$ process for the LHC Run 2+3 and HL-LHC luminosity, based on which the total events are sliced into 1000 pseudo experiments. Then using the theoretical framework, all the coefficients of the spin density matrix  is determined by the angular distribution of the final lepton pairs coming form $Z$ boson decay. Entanglement quantifier, the lower bound of the concurrence, is measured using these pseudo experiments. The significance of this quantity can reach to 2$\sigma$ with combined LHC Run 2+3 data and 3.75 $\sigma$ for HL-LHC data, making it possible to measure quantum entanglement between two massive gauge boson pairs in an actual LHC experiment.

\begin{acknowledgments}
This work is supported in part by the National Natural Science Foundation of China under Grants No. 12325504, No. 12150005, and No. 12075004. 
\end{acknowledgments}

\appendix
\section{The functions \(\mathfrak{q}^{n}_\pm\) and \(\mathfrak{p}^n_\pm\)}

The functions \(\mathfrak{q}^{n}_\pm\) and \(\mathfrak{p}\) are defined as follows according to ~\cite{fabbrichesi2023bell,ashby2023quantum}:
\begin{align}
    \mathfrak{q}_\pm^1 &= \frac{1}{\sqrt{2}} \sin \theta^\pm \left(\cos\theta^{\pm}\pm 1\right)\cos\phi^\pm, \nonumber\\
    \mathfrak{q}_\pm^2 &= \frac{1}{\sqrt{2}} \sin \theta^\pm \left(\cos\theta^{\pm}\pm 1\right)\sin\phi^\pm, \nonumber\\
    \mathfrak{q}_\pm^3 &= \frac{1}{8} \left(1\pm4\cos\theta^\pm+3\cos2\theta^\pm\right), \nonumber\\
    \mathfrak{q}_\pm^4 &= \frac{1}{2}\sin^2\theta^\pm \cos 2 \phi^\pm, \nonumber\\
    \mathfrak{q}_\pm^5 &= \frac{1}{2}\sin^2\theta^\pm \sin 2 \phi^\pm, \nonumber\\
    \mathfrak{q}_\pm^6 &= \frac{1}{\sqrt{2}} \sin\theta^\pm \left(-\cos\theta^\pm\pm1\right)\cos\phi^{\pm}, \nonumber\\
    \mathfrak{q}_\pm^7 &= \frac{1}{\sqrt{2}} \sin\theta^\pm \left(-\cos\theta^\pm\pm1\right)\sin\phi^{\pm}, \nonumber\\
    \mathfrak{q}_\pm^8 &= \frac{1}{8\sqrt{3}}\left(-1\pm12\cos\theta^\pm-3\cos2\theta^\pm\right),
    \label{eq:Qsymbol}
\end{align}
where \(\theta^\pm\) and \(\phi^\pm\) are the polar and azimuthal angles of the decay lepton \(l^\pm\) in the \(Z\) boson rest frame, respectively. The function \(\mathfrak{p}\) is defined as:
\begin{align}
    \mathfrak{p}_\pm^1 &= \sqrt{2} \sin\theta^\pm \left(5\cos\theta^\pm\pm1\right)\cos\phi^\pm, \nonumber\\
    \mathfrak{p}_\pm^2 &= \sqrt{2} \sin\theta^\pm \left(5\cos\theta^\pm\pm1\right)\sin\phi^\pm, \nonumber\\
    \mathfrak{p}_\pm^3 &= \frac{1}{4} \left(5\pm4\cos\theta^\pm+15\cos2\theta^\pm\right), \nonumber\\
    \mathfrak{p}_\pm^4 &= 5\sin^2\theta^\pm \cos 2 \phi^\pm, \nonumber\\
    \mathfrak{p}_\pm^5 &= 5\sin^2\theta^\pm \sin 2 \phi^\pm, \nonumber\\
    \mathfrak{p}_\pm^6 &= \sqrt{2} \sin\theta^\pm \left(-5\cos\theta^\pm\pm1\right)\cos\phi^{\pm}, \nonumber\\
    \mathfrak{p}_\pm^7 &= \sqrt{2} \sin\theta^\pm \left(-5\cos\theta^\pm\pm1\right)\sin\phi^{\pm}, \nonumber\\
    \mathfrak{p}_\pm^8 &= \frac{1}{4\sqrt{3}}\left(-5\pm12\cos\theta^\pm-15\cos2\theta^\pm\right).
    \label{eq:Psymbol}
\end{align}
The matrix \(\mathfrak{a}_m^n\)is defined as:
\begin{align}
    \mathfrak{a}_m^n = \frac{1}{g_L^2-g_R^2} \begin{bmatrix}
        g_R^2 & 0 & 0 & 0 & 0 & g_L^2 & 0 & 0 \\
        0 & g_R^2 & 0 & 0 & 0  & 0 & g_L^2 & 0 \\
        0 & 0 & g_R^2-\frac{1}{2}g_L^2 & 0 & 0  & 0 & 0 & \frac{\sqrt{3}}{2} g_L^2 \\
        0 & 0 & 0 & g_R^2-g_L^2 & 0 & 0 & 0 & 0 \\
        0 & 0 & 0 & 0 & g_R^2-g_L^2 & 0 & 0 & 0 \\
        g_L^2 & 0 & 0 & 0 & 0 & g_R^2 & 0 & 0 \\
        0 & g_L^2 & 0 & 0 & 0 & 0 & g_R^2 & 0 \\
        0 & 0 & \frac{\sqrt{3}}{2} g_L^2 & 0 & 0 & 0 & 0 & g_L^2-\frac{1}{2}g_R^2 
    \end{bmatrix}
    \label{eq:amatrix}
\end{align}
\section{Gell-Mann matrices}
The \(3\times3\) Gell-Mann matrices are defined as:
\begin{align}
\begin{matrix}
    \mathbb{1}=\begin{bmatrix}1 & 0 & 0 \\ 0 & 1 & 0 \\ 0 & 0 & 1 \end{bmatrix}, &
    T^1 = \begin{bmatrix} 0 & 1 & 0 \\ 1 & 0 & 0 \\ 0 & 0 & 0 \end{bmatrix}, &
    T^2 = \begin{bmatrix} 0 & -i & 0 \\ i & 0 & 0 \\ 0 & 0 & 0 \end{bmatrix}, \nonumber\\
    T^3 = \begin{bmatrix} 1 & 0 & 0 \\ 0 & -1 & 0 \\ 0 & 0 & 0 \end{bmatrix}, &
    T^4 = \begin{bmatrix} 0 & 0 & 1 \\ 0 & 0 & 0 \\ 1 & 0 & 0 \end{bmatrix}, &
    T^5 = \begin{bmatrix} 0 & 0 & -i \\ 0 & 0 & 0 \\ i & 0 & 0 \end{bmatrix}, \nonumber\\
    T^6 = \begin{bmatrix} 0 & 0 & 0 \\ 0 & 0 & 1 \\ 0 & 1 & 0 \end{bmatrix}, &
    T^7 = \begin{bmatrix} 0 & 0 & 0 \\ 0 & 0 & -i \\ 0 & i & 0 \end{bmatrix}, &
    T^8 = \frac{1}{\sqrt{3}} \begin{bmatrix} 1 & 0 & 0 \\ 0 & 1 & 0 \\ 0 & 0 & -2 \end{bmatrix}.
\end{matrix}
\end{align}

\begin{acknowledgments}
This work is supported by National Natural Science Foundation of China (NSFC) under Grants No.~12325504 and No.~120611410, and by MOST under grant No.~2023YFA1605800.
\end{acknowledgments}

\bibliographystyle{ieeetr}
\bibliography{reference}

\end{document}